\documentclass[twocolumn,showpacs,amsmath,amssymb]{revtex4}
\usepackage{graphicx}
\usepackage{dcolumn}
\usepackage{bm}
\usepackage{amsmath}
\usepackage{amsfonts}
\usepackage{booktabs,multirow}
\usepackage{multirow}
\usepackage{epstopdf}
\usepackage{caption}
\usepackage{subfigure}
\begin{document}

\title{Blocking the Hawking Radiation}
\author{Martin {\sc Autzen}}\email{autzen@cp3-origins.net} \author{Chris {\sc Kouvaris}}\email{kouvaris@cp3.sdu.dk}
\affiliation{$\text{CP}^3$-Origins, University of Southern Denmark, Campusvej 55, Odense 5230, Denmark}

\begin{abstract}
Some severe constraints on asymmetric dark matter are based on the scenario that certain types of WIMPs can form mini-black holes inside neutron stars that can lead to their destruction. A crucial element for the realization of this scenario is that the black hole grows after its formation (and eventually destroys the star) instead of evaporating. The fate of the black hole is dictated by the two opposite mechanics i.e. accretion of nuclear matter from the center of the star and Hawking radiation that tends to decrease the mass of the black hole. We study how the assumptions for the accretion rate can in fact affect the critical mass beyond which a black hole always grows. We also study to what extent degenerate nuclear matter can impede Hawking radiation due to the fact that emitted particles can be Pauli blocked at the core of the star. \\[.1cm] {\footnotesize \it Preprint: CP$^3$-Origins-2014-006
  DNRF90 \& DIAS-2014-6.}
\end{abstract}


\maketitle 

\section{Introduction}
Observations of old compact stars have been used in order to impose constraints on  specific types of (a)symmetric dark matter~\cite{Goldman:1989nd,Kouvaris:2007ay,Bertone:2007ae,Kouvaris:2010vv,deLavallaz:2010wp,Kouvaris:2011fi,McDermott:2011jp,Guver:2012ba,Bell:2013xk,Bramante:2013hn,Bramante:2013nma,
Kouvaris:2011gb,Capela:2012jz,Capela:2013yf,Kouvaris:2010jy,Fan:2012qy} or to predict new effects~\cite{PerezGarcia:2010ap,PerezGarcia:2011hh}.  A set of the derived constraints is based on the fact that asymmetric dark matter can be trapped inside compact stars and under certain conditions, the WIMP population might collapse forming a black hole at the center of the star which eventually can consume it. The observation of old neutron stars (as it is well established) can therefore eliminate specific dark matter candidates,  because their existence would have implied the destruction of these stars by WIMP-generated black holes inside the stars. However, in order to consider these constraints seriously, one has to ensure that all the stages that lead to the destruction of the star take place and that the destruction does not happen in time scales larger than billions of years. The last is important because in principle black holes might exist inside stars but due to small accretion rates, they could potentially not have a visible effect yet.  At this point one should recall that in the case of repulsively interacting-bosons, the mass that leads to a black hole formation is~\cite{Kouvaris:2011fi}
\begin{equation}
M_c=\frac{2}{\pi}\frac{M_{\rm pl}^2}{m}  \sqrt{1+\frac{M_{\rm pl}^2}{4\sqrt{\pi}m}\sigma^{1/2}}, \label{chandra}
\end{equation}
where $M_{\rm pl}$ and $m$ are the Planck mass and the WIMP mass respectively and $\sigma$ is a repulsive WIMP-WIMP interaction cross section modeled via a $\phi^4$ type of interaction. In the absence of self-interactions ($\sigma=0$) it is easy to deduce the above formula (up to a numerical factor of order one) by demanding the self-gravitation potential energy of the WIMP population to be larger than the relativistic kinetic energy coming from the uncertainty principle. The kinetic energy is $ \hbar /r$ and therefore the criterion for collapse becomes
\begin{equation}
\frac{\hbar}{r}<\frac{GMm}{r},
\end{equation}
that leads to Eq.~(\ref{chandra}). 

However, after the formation, the expansion of the black hole depends on two competing mechanisms: accretion of the surrounding matter at the core of the star that obviously tends to increase the mass of the black hole and Hawking radiation that tends to reduce the mass of the black hole. 
The rate of change for the black hole mass is 
\begin{equation}
\frac{dM}{dt}=CM^2-\frac{f}{G^2M^2}. \label{evol}
\end{equation}
In the case of spherical accretion in the hydrodynamic limit (Bondi accretion) $C=4\pi \lambda_s \rho_c G^2/c_s^3$, where $\lambda_s$ is a coefficient of order one, $\rho_c$ is the matter density at the core of the star, $M$ is the mass of the black hole, and $c_s$ is the speed of sound for the accreting matter. $f$ is a dimensionless number giving the power radiated away from Hawking radiation and it depends on the number of modes participating (how many different species of particles are emitted) and how fast the black hole rotates.
Since accretion scales as $M^2$ and Hawking radiation as $M^{-2}$, there is a critical value $M_{\rm crit}=(f/G^2C)^{1/4}=m_{\rm pl}(f/C)^{1/4}$ above which the black hole grows and below which it is doomed to evaporate. Therefore if $M_c<M_{\rm crit}$, the black hole evaporates, while in the opposite case it grows. Due to the dependence of $M_c$ on $m$  in Eq.~(\ref{chandra}), the existence of the $M_{\rm crit}$ sets an upper bound $m_{\rm upp}$ on the WIMP mass upon the constraints can be potentially applied. For WIMPs with $m>m_{\rm upp}$, the mass of the formed black hole $M_c$ is smaller than $M_{\rm crit}$ and therefore the black hole eventually evaporates, so no constraints can be drawn.

The process of the potential destruction of the star has the following stages: accretion of WIMPs onto the stars, thermalization with the surrounding nuclear matter and concentration at the core of the star, Bose-Einstein Condensate formation, self-gravitation, loss of energy of the WIMP sphere, formation of the black hole, and expansion of the black hole. If any of these stages does not take place, no destruction of the star happens and the constraints are invalid. Issues regarding the thermalization time scale of the WIMPs inside the star have been addressed in~\cite{Bertoni:2013bsa}. The effect of rotation of the neutron star on the rate of expansion of the black hole was addressed in~\cite{Kouvaris:2013kra}. One should mention that in the case of non-interacting bosonic WIMP with masses above $\sim 10$ TeV, the time order of BEC formation and self-gravitation is  reversed. However, as it was pointed out in~\cite{Kouvaris:2012dz}, although the self-gravitating WIMP population might have a mass larger than $M_{\rm crit}$, the population does not collapse altogether, but rather  black holes of $M_c$ are formed one after the other where every time the black hole evaporates before the next one forms. This means that no constraints can be applied in this case.   

The $m_{\rm upp}$ in the absence of self-interactions was estimated in~\cite{Kouvaris:2011fi} to be around 16 GeV assuming that  only photons are emitted via Hawking radiation.
In this paper we estimate more precisely the critical black hole mass $M_{\rm crit}$ and the upper bound on the WIMP mass that the constraints can be applied taking into consideration two things: The first one has to do with the power of Hawking radiation. As it is known, a black hole of temperature $T$ should roughly emit all elementary particles with mass lower than T.
However, since the newly formed black hole is immersed in degenerate matter, some of the modes can be (partially) blocked. 
The temperature of a newly formed  black hole is 
\begin{equation}
T=\frac{1}{8\pi GM_c}=\frac{m}{16}. \label{temp}
\end{equation}
where we used Eq.~(\ref{chandra}) (in the case where self-interactions are absent) in the final part. For a WIMP mass of a few GeV which is roughly the upper bound on $m$ deduced previously, the temperature is just below GeV. Given that at the core of a neutron star, the baryon chemical potential is of the order of GeV, some of the modes might be partially or fully blocked due to degeneracy. This can potentially reduce significantly Hawking radiation, thus increasing the upper bound for $m$ (where constraints can be applied). On the other hand, the newly formed black hole can have a Schwarzschild radius that can be small compared to nucleon sizes. 

Using $r_s=2GM$ and Eq.~(\ref{chandra}) for example in the absence of self-interactions, one gets 
\begin{equation}
r_s=2GM_c=\frac{4}{\pi m}\simeq 2.5 \times 10^{-14} \left ( \frac{{\rm GeV}}{m} \right ) {\rm  cm}, \label{r_s}
\end{equation}
 which is smaller than the proton size already for $m=1$ GeV. Therefore, the size of the black hole can be smaller than the size of the particles it accretes. From this point of view, it is not clear at all, if the conditions for Bondi accretion are fulfilled. The Bondi accretion solution is derived upon assuming that the accreted matter behaves as a smooth fluid. However, it is not clear if the hydrodynamic limit is satisfied. We are going to check how the upper limit for the WIMP mass where the existing constraints can still apply is modified if instead of a Bondi accretion we simply assume a geometric cross section for the black hole-matter collision.

\section{Accretion}
As we mentioned, it is not clear if the assumptions for Bondi accretion are fulfilled for small black holes where the accreted particles are much larger than the size of the black hole. However, no matter whether we have Bondi accretion or not, the black hole can definitely accrete matter with a geometric cross section. This should be the lower possible accretion rate for any black hole. The core of a neutron star can be made of quarks or nucleons depending on the core density, but in any case, matter is degenerate. Let us assume that we have a particle (either quark or nucleon) with some Fermi momentum $k_F$ and a black hole with a mass $M$ situated at the core of the star. One can estimate what is the rate of accretion based on a geometric cross section i.e. we can estimate the flux of particles (and energy) crossing the event horizon and therefore contributing to the growth of the black hole. The energy density in a given volume is simply $du=2Ed^3k/(2 \pi)^3$ where $E$ is the energy of a particle with momentum $k$. The factor of 2 is due to the fact that there are two spin states since we are going to consider fermions with spin $1/2$. The energy flux i.e. energy per surface area per time of particles passing through the event horizon is 
\begin{equation}
F=\frac{2}{(2\pi)^3}\int Ev \cos \theta f(k) k^2 dk d(\cos \theta) d\phi,
\end{equation}
where $f(k)$ is the Fermi-Dirac statistical distribution, and $\theta$ is the angle of the velocity of a particle with respect to the line that connects the position of the particle and the center of the black hole. The integration over $\phi$ gives trivially $2\pi$, while we integrate over $\theta$ from 0 up to $\pi/2$ (so we count the particles that have velocity components that can make the particle cross the event horizon). Keeping in mind that $v=k/E$, and multiplying with the total surface of the black hole $4 \pi r_s^2=16 \pi G^2M^2$ we get the total energy flux per time to lowest approximation of $T/\mu$ ($T$ being the temperature)
\begin{equation}
\frac{dE}{dt}=\frac{1}{\pi}G^2M^2k_F^4.
\end{equation}
This lower bound on the accretion of particles should be contrasted to the Bondi accretion rate (i.e. $dE/dt=4\pi \lambda_s \rho_c G^2/c_s^3$). If the non-Bondi accretion rate is much smaller than the Bondi rate, say by a factor $\epsilon$, $m_{\rm upp}$ will lower by a factor $\epsilon^{1/4}$.

\section{Hawking Radiation}
As we mentioned, any elementary particle with a mass smaller than the temperature of the black hole should be able to be emitted via the Hawking mechanism. The Hawking radiation spectrum is not exactly identical to the black body one. Page in his seminal paper~\cite{Page:1976df} estimated the number and energy emission rates for fermions and bosons. The overall energy rate for the emission is
\begin{equation}
\frac{dM}{dt}=-(n_f f_f +n_b f_b + n_s f_s +n_2 f_2) \frac{1}{G^2M^2}, \label{block}
\end{equation}
where $n_i$ denotes the number of particle species multiplied by the number of helicity states  for $i=(f,b,s,2)$ i.e. respectively fermion, boson, scalar and spin 2
particles. There is only one scalar (Higgs boson), and one spin-2 particle i.e. the graviton. The values of the corresponding numbers $f_i$ can be read off Table I of~\cite{Page:1976df} if one sums the contributions of all the different modes i.e. modes carrying angular momentum $l$ multiplied by a factor $2(2l+1)$. It is understood that each quark comes with a multiplicity of three due to color and that the total number of bosons is nine (i.e. eight gluons and the photons).  
Now we shall estimate which species of particles contribute to Hawking radiation and to what extent these modes are Pauli blocked.

\subsection{ Quark Blocking}
We are going to consider the scenario where the center of the neutron star is made of noninteracting quark matter. Once the matter density becomes sufficiently higher than the nuclear density, protons and neutrons at the center of the neutron star have overlapping wavefunctions and therefore quarks are the appropriate degrees of freedom 
to consider. For a  typical quark chemical potential  $\mu \sim 500$ MeV, three quarks are present i.e. $u$, $d$, and $s$.  This quark matter has to satisfy overall electric neutrality because a net nonzero electric charge will cause
huge Coulomb energy costs. In principle equal numbers of $u$, $d$, and $s$ should automatically satisfy electric neutrality. However, since the strange quark mass $m_s \sim 120$ MeV is much larger than the corresponding $m_u$ and $m_d$,  and comparable to $\mu$, strange quarks are less abundant and the presence of electrons is needed to ensure the neutrality. Weak equilibrium relates the chemical potentials of the three quarks and electrons as:
 \begin{eqnarray} \mu_u & = &\mu-\frac{2}{3} \mu_e \nonumber \\ 
  \mu_d & = &\mu+\frac{1}{3} \mu_e  \nonumber  \\
 \mu_s & = &\mu+\frac{1}{3} \mu_e, \label{eq1}
\end{eqnarray}
 where the index obviously refers to the corresponding particle. Assuming zero mass for $u$, and $d$ (which is a good approximation given they are much smaller than $m_s$ and $\mu$), quarks fill all the states (in the momentum space) up to
 the Fermi surfaces:
  \begin{eqnarray}
p_F^u & = & \mu_u \nonumber \\
   p_F^d & = & \mu_d \nonumber \\
   p_F^s & = & \sqrt{\mu_s^2-m_s^2}. 
   \end{eqnarray}
   
   The free energy of the system is
   \begin{widetext}
    \begin{equation}
    \Omega =  \frac{3}{\pi^2} \int_0^{p_F^u} (p-\mu_u) p^2 dp + \frac{3}{\pi^2} \int_0^{p_F^d} (p-\mu_d) p^2 dp  + \frac{3}{\pi^2} \int_0^{p_F^s} (\sqrt{p^2+m^2}-\mu_s) p^2 dp + \frac{1}{\pi^2} \int_0^{p_F^e} (p-\mu_e) p^2 dp. \end{equation}
   \end{widetext} The factor of 3 in the quarks contributions is related to the number of colors. The minimization of the free energy with respect to $\mu_e$ imposes the electric neutrality. To lowest order in $m_s$, the free energy minimizes for~\cite{Alford:2002kj}  
   \begin{equation} \mu_e=\frac{m_s^2}{4\mu}. \end{equation}  Eq.~(\ref{eq1}) now reads
   
    \begin{eqnarray}
     p_F^u & = &\mu-\frac{m_s^2}{6\mu}  \nonumber \\ 
     p_F^d & = &\mu+\frac{m_s^2}{12\mu}  \nonumber  \\
     p_F^s & = &\mu-\frac{5m_s^2}{12\mu}.  \label{fermi_m}
  \end{eqnarray}
  One can observe that $d$ quarks have the highest Fermi momentum, followed by $u$, and $s$. It can also be seen that there is equal difference in the splitting of the Fermi momenta of the three quarks i.e. 
 $p_F^d-p_F^u=p_F^u-p_F^s=m_s^2/4\mu$. The fact that $m_s^2/\mu$ is not much smaller than $\mu$ makes it more difficult for $s$ quarks to be present and therefore nonzero density of electrons is needed to preserve the
 neutrality of the system. In this case we have ignored strong interactions and therefore quarks of the same flavor but different color have the same Fermi momentum. Turning on the strong interactions can split the Fermi momenta even among 
 quarks of the same flavor but different color and can lead to color superconducting states (see \cite{Rajagopal:2000wf,Alford:2001dt} for reviews). In this case apart from electric neutrality, one should impose in addition color neutrality   (which is automatically satisfied in the case of
 noninteracting quarks that we studied). Although some excitation above the Fermi surface might acquire an energy gap, no species of quarks can have a  momentum considerably smaller than $\mu$, which is what we need 
  for our argument with respect to Hawking radiation.
\begin{figure}[h]
\label{fig:FermiMode}
\centering
\includegraphics[scale=.7]{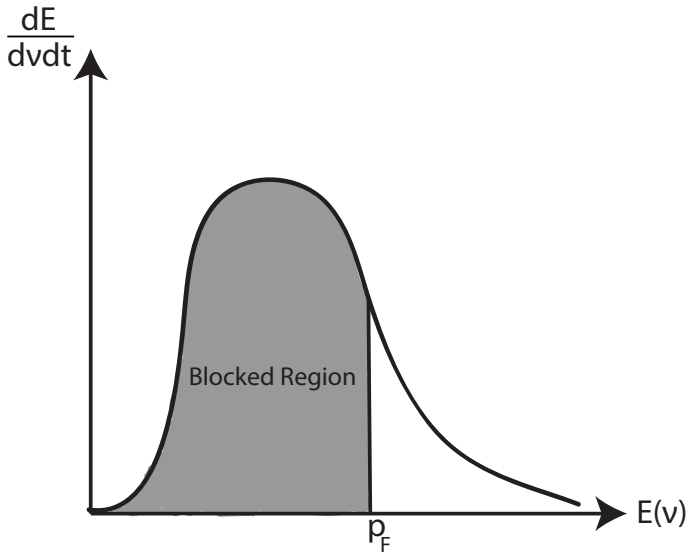}
\caption{Blocking of Hawking radiation due to degeneracy for modes with $p<p_F$.}
\end{figure}

As we mentioned, if one assumes Bondi accretion and Hawking radiation with only photons, $m_{\rm upp}$ in the absence of self-interactions is of the order of $\sim10$ GeV. Using Eq.~(\ref{temp}), this means that the temperature of the black hole at formation is $\sim 600$ MeV. Hawking radiation is usually envisioned as a process where a pair of particle-antiparticle is created close to the event horizon of the black hole where one of them tunnels inside the black hole whereas the second particle escapes. In principle, as long as the temperature allows it, any pair of particles can be created. 
The characteristic size of a newly formed black hole made of  WIMPs (see Eq.~(\ref{r_s}))  is much smaller than the size of composite objects such as pions, protons, neutrons etc. A newly formed black hole made of 10 GeV asymmetric bosonic WIMPs has a Schwarzschild radius of $2.5 \times 10^{-15}$ cm, which is almost two orders of magnitude smaller than a typical nuclear size. Therefore it is obviously more appropriate to consider pair production of elementary particles rather than composite hadrons. With a temperature of 600 MeV or less, $W^{\pm}$, and $Z$ bosons, $t$, $b$, and $c$ quarks, as well as $\tau$ leptons cannot be pair-produced from the vacuum. The particles that can in principle be produced are $u$, $d$, and $s$ quarks, as well as all the neutrinos, $e$, and $\mu$ from the leptons, and photons, gluons and gravitons. However, as we are going to demonstrate, some of the particles are (partially) blocked due to degeneracy and contribute less to the emission power of Hawking radiation.

In order to show this, we can use Eq.~(\ref{fermi_m}) for a typical $m_s=120$ MeV and $\mu=500$ MeV. The Fermi momenta for $d$, $u$, $s$, and $e$ are respectively 502, 495, 488, and 7 MeV. Let us now move to the basics of spherical accretion of matter into a black hole (Bondi accretion). The reader can review Bondi accretion in e.g.~\cite{Shapiro:1983du}. If one assumes an adiabatic flow of matter in the black hole, the process will be characterized by a polytropic equation of state

\begin{equation} P= K n^{\Gamma}  \end{equation}
 with $K$ and $\Gamma$ constants, and $n$ being the number density of particles.  In Bondi accretion  matter flows towards the black hole with a subsonic velocity until it reaches a distance from  the center of the black hole
\begin{equation}
r_B= \frac{5-3 \Gamma}{4} \frac{GM}{c_s^2}, \label{rs}
\end{equation}  where $c_s$ is the sound speed of infalling matter. When matter passes this point (the Bondi radius $r_B$), it moves towards the black hole supersonically.
Assuming relativistic matter (which is certainly true for quarks with $\mu=500$ MeV), $\Gamma=4/3$. Now we need an estimate for the sound speed $c_s$. The sound speed in a neutron star with typical densities of nuclear matter has been estimated to 0.17$c$ \cite{Kouvaris:2011fi}. However, relativistic non-interacting quark matter would have a sound speed of $1/\sqrt{3}$. It is expected that quark interactions will change the sound speed. In fact studies using a modified version of the MIT Bag model show that  the sound speed can be within 0.2-0.4$c$ for densities larger by a factor of a few with respect to the nuclear density~\cite{Wen:2009zza}. However due to the uncertainty in the matter density inside the star, we are going to consider the full range of $c_s$ from 0.17$c$ to $1/\sqrt{3}\simeq 0.58c$.

In Bondi accretion, the density of infalling matter below the Bondi radius is~\cite{Shapiro:1983du} 
\begin{equation}
\frac{n(r)}{n_{\infty}} \simeq \frac{ \lambda_s}{ \sqrt{2}} \left ( \frac{GM}{c_s^2r} \right )^{3/2},
\end{equation}
where $n_{\infty}$ is the particle density at  asymptotically large distances (much larger than the Bondi radius). Close to the event horizon which is the region where Hawking radiation is produced, 
\begin{equation}
\frac{n_h}{n_{\infty}} \simeq \frac{ \lambda_s}{ 4c_s^3}. \label{scale}
\end{equation}
For $c_s=0.17c$, the ratio is $\sim 36$ where for $c_s=0.58c$ the ratio is close to one. Since number densities scale as $\sim k_F^3$, an increase of 36 in the density would correspond to an increase of $\sim 36^{1/3}=3.3$ in the Fermi momentum in the case of $c_s=0.17c$. 

Two comments are in order here. The first one has to do with weak equilibration. Matter flows towards the black hole with an increasing density. In the case of chemical equilibrium, larger density means larger chemical potential and therefore one could expect that the Fermi momenta of $u$, $d$, $s$, and $e$ might not be simply rescaled by the same factor, but rather might be determined by the minimization of the free energy we performed earlier with an appropriate chemical potential larger than 500 MeV. However, this is not the case.
The cross section for weak equilibration scales as $\sim G_F^2p_FT$ (where $G_F$ is the Fermi constant, $p_F$ the Fermi momentum and $T$ the temperature. The time scale for weak equilibration should be $1/(G_F^2p_FTn)$ where $n$ is the number density of  quarks that are not Pauli blocked. In any case this time scale is much longer than the dynamic time scale $(R^3/GM)^{1/2}$ and therefore we can safely assume that all components scale by the same factor since weak interactions do not have the time to  convert for example electrons and up quarks to down quarks.

The second remark has to do with  a potential deformation of the Fermi surface. At large distances from the Bondi radius, the Fermi surface is spherical. However, in Bondi accretion matter starts flowing with almost a free fall velocity below the Bondi radius. This means that all particles acquire a boost in the radial direction. Since this happens for all in-falling particles, one could expect that although higher momenta are being occupied, low momentum modes become available and thus unblocked. However one should not worry for this issue for three reasons. The first one has to do with the fact that although the Fermi surface might be deformed in the radial direction, all other directions remain with the old asymptotic values. Therefore the phase space for the emission of unblocked low momentum particles is confined strictly along the radial direction. The second reason has to do simply with the fact that the power output is small for low momenta. However, these modes are problematic even for another reason. The emitted fermion spectrum peaks at energies $\sim 4.5T_H$~\cite{Page:1976df}. The thermal De Broglie wavelength of such a particle is $\lambda_d=\pi^{2/3}/(4.5T_H)$. If one assumes that the mass of the black hole is the critical one given by (\ref{chandra}), then for example in the abcence of self-interactions $\lambda_d=1.5 m \times 10^{-13}~\text{cm}$, where $m$ is the WIMP mass in GeV. 
  Although this length is derived under the assumption of massless particles, for all particles under consideration, it is a good approximation. This characteristic De Broglie length is much larger than the Bondi radius (for all WIMPs in the keV to $\sim10$ GeV range).The flow velocity of matter at distances $r>>r_s$ is~\cite{Shapiro:1983du} 
\begin{equation} \frac{u}{c_s}\simeq \lambda_s \left (\frac{GM}{c_s^2} \right )^2r^{-2}. \label{radius} 
\end{equation}   
  The flow velocity at distances $r\sim \lambda_d$ becomes a tiny fraction of 
the sound speed. Therefore, we can safely assume that at distances $r\sim \lambda_d$, the Fermi surface is spherical and it has not been disfigured by the acceleration towards the black hole.  Therefore, even if low momentum particles are emitted, they will be blocked since at distances of the order of their De Broglie length, the Fermi surface is completely spherical.

Hawking emission of fermionic particles can in principle be Pauli blocked if the particle has a momentum below the Fermi momentum of that particle (if it is present outside the black hole). The blocking is illustrated in Fig.~1.
 As we mentioned earlier, typical quark matter chemical potential $\mu=500$ MeV will lead to Fermi momenta for $d$, $u$, $s$, and $e$  of 502, 495, 488, and 7 MeV respectively. We also mentioned that
 for spin $1/2$ particles, the spectrum peaks at $\sim 4.5T_H=0.28m$. Here we should note that although from now on  we focus in the case where self-interactions among WIMPs are absent, our arguments do not depend on this. The $M_{\rm crit}$ that we will derive after taking into account the blocking effect will be valid whether or not there are WIMPs self-interactions. For heavy black holes the corresponding temperature is small and the blocking effect will be significant. However for heavy black holes already accretion wins over Hawking radiation and therefore the black hole growth is not in danger. On the other hand, lighter black holes that have potentially large temperatures are of interest for our study since it is this regime where the black hole growth is in danger. One can see for example that  in the case of a black hole formed from a 5 GeV WIMP, the spin $1/2$ spectrum peaks at $\sim 1.4$ GeV which is much 
 larger than any of the Fermi momenta of $d$, $u$, $s$, and $e$. However as we pointed out earlier, for $c_s=0.17c$, close to the horizon, the Fermi momentum increases by a factor of 3.3. This means that a Fermi momentum of $\sim 500$ MeV will be boosted to $\sim 1.65>1.4$ GeV. Therefore a significant part of the spectrum of Hawking radiation to quarks can be significantly blocked. It is understood that the blocking becomes less severe as $c_s$ increases. Since for $c_s=0.58c$ there is not any significant change between the asymptotic density and the horizon one, only a small part (of low momentum) of the spectrum will be blocked.  
  
 \subsection{Gluon Blocking} 
  
  The situation regarding the gluons is different. Although gluons are bosons and do not form a Fermi surface, the creation of a pair of gluons where one is tunneled inside the black hole and the other escapes can also be problematic. Let us imagine a gluon emitted with an energy at the peak of the spectrum. For spin 1 particles, the peak is  higher compared to the spin $1/2$ particles. The
  peak is at $\sim 5.8 T_H=0.36m$ (if we use Eq.~(\ref{temp})). This means that in the case that the black hole is formed by a 5 GeV WIMP, the gluon would have energy 1.8 GeV. 
  A gluon can be always thought to carry $\bar{q}q$ quantum numbers, where the $q$ quark and the $\bar{q}$ antiquark carry different colors. The two gluons will create a color flux tube of two $\bar{q}q$ extending from the black hole to the position of the escaping gluon. The color flux tube can break once it becomes energetically favorable to create two intermediate $\bar{q}q$ that will hadronize the escaping gluon and the gluon inside the black hole. This hadronization can potentially produce two escaping pions that being with a mass 140 MeV, they pass the energy threshold i.e. $2 \times 140~\text{MeV}< 1.87~\text{GeV}$ (in the case of a 5 GeV WIMP). Although this will be a possible scenario in a black hole emitting gluons in the vacuum, the situation changes due to the degeneracy of matter. In order to create the intermediate $\bar{q}q$ pairs to hadronize the escaping gluon, one should provide energy twice the Fermi energy of the degenerate quarks which we estimated to be $\sim 500$ MeV (if $\mu =500$ MeV). For each of the two $\bar{q}q$
 pairs, $q$ and $\bar{q}$ have to be created from the vacuum back-to-back. This means that for each pair the energy required is twice the Fermi momentum of the corresponding quark (i.e. $\sim 2 \times 500$ GeV). Therefore the total required energy for hadronization is 2 GeV which is larger than the $1.8$ GeV which is the energy at the peak of the photon spectrum. 
   In such a case the escaping gluon cannot hadronize, at least until it can get to a region with lower quark chemical potential. However, this is practically impossible since the energy cost for extending the color flux tube without breaking it for more than a few $fm$ will be tremendous. The situation is depicted in Fig.~2. The blocking of gluons is more prominent also due to the quark compression in Bondi accretion we mentioned in the previous subsection. Since quarks become denser close to the horizon, we found that their ``effective'' Fermi momentum increases by a factor up to 3.3 (for $c_s=0.17$). In that case the required energy for hadronization will be 3.3 times the 2 GeV we mentioned earlier, making the blocking of gluons almost absolute since the peak of the spectrum is at 1.8 GeV and the blocking up to 6.6 GeV.
  
   \begin{figure}[ht]
	\centering
	\begin{subfigure}[]
		\centering
		\includegraphics[scale=0.7]{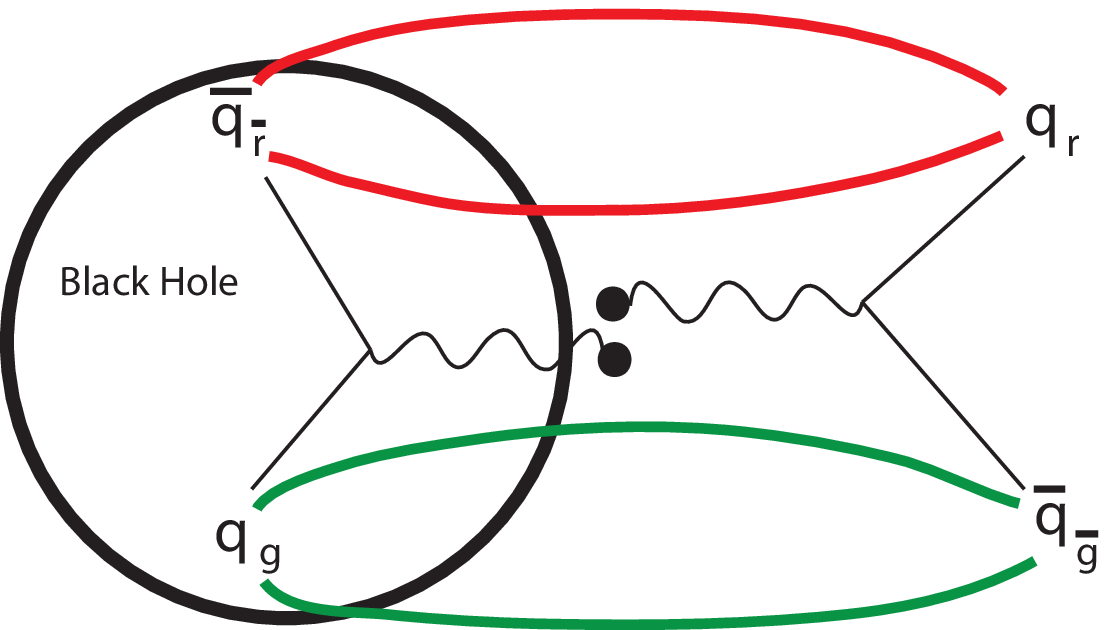}
		\label{fig:ColorFluxTube}
	\end{subfigure}
	\quad
	\begin{subfigure}[]
			\centering
		\includegraphics[scale=0.4]{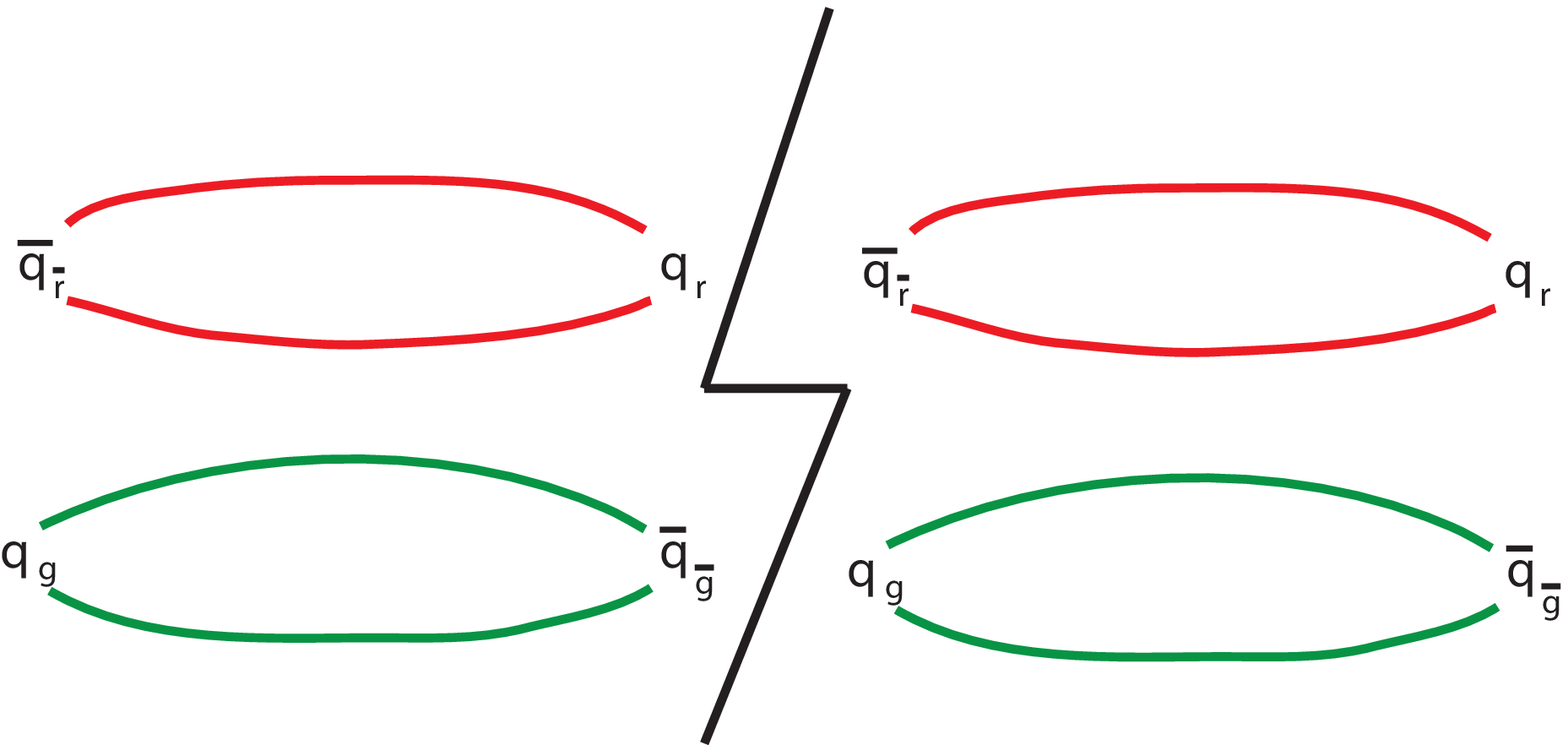}
		\label{fig:ColorFluxBreaking}
	\end{subfigure}
	\caption{These figures illustrate how an emitted gluon can be neutralized by two $\bar{q}q$ pairs created from the vacuum.}
\end{figure}
 In principle there is one other way that gluons can color-neutralize apart from creating quark-antiquark pairs from the vacuum. This can happen if the gluon is neutralized by another gluon emitted by the black hole in a later time. This way it would not be necessary to create  $\bar{q}q$ from the vacuum. This is possible if the time between two successive emissions is shorter than the inverse of the frequency associated with the De Broglie wavelength of the emitted gluon. A rough estimate for the number of particles emitted per time for a light black hole is~\cite{MacGibbon:1990zk}
  \begin{equation}
 \frac{dN}{dt}=10^{-2}\frac{1}{GM}.
 \end{equation}
The mean time between emissions is 
\begin{equation}
\Delta t = 100 GM=\frac{200}{\pi m}>>\lambda_d=\frac{2\pi}{0.37m},
\end{equation}
where we used Eq.~(\ref{chandra}) (with $\sigma=0$) for the mass of the black hole and the De Broglie wavelength that corresponds to the peak energy of the gluon spectrum. This time is also longer from the time it takes for the gluon to stretch the colored string to distances of $fm$. This shows that the time for the successive emission of two gluons is much longer than the time it takes for the gluon to feel the blocking of quarks and therefore the gluon cannot hadronize. 

\section{Results}
In this section we present the results regarding the amount of blocking of the different modes under consideration. We use Eq.~(\ref{evol}) to find the critical mass $M_{\rm crit}$ above which the black hole grows in the environment of a neutron star core. We do this by using the more accurate expression of Eq.~(\ref{block}) for the Hawking radiation, blocking the emitted modes with momentum below the Fermi momentum of each particular species.

In Table I we show $M_{\rm crit}$  for the case of Bondi accretion for the three distinctive scenaria of no blocking, simple blocking and boosted blocking. By  simple blocking we mean that we considered blocking of the modes up to the  Fermi momentum of each species at asymptotically large distances. Boosted blocking refers to the enhanced Fermi momentum due to the fact that accreting particles become denser close to the horizon as we explained below Eq.~(\ref{scale}). 
 As one can see the effect of blocking is small $\sim 10\%$. In addition in Table I we show the corresponding maximum WIMP mass (converted from $M$ via Eq.~(\ref{chandra}) for $\sigma=0$) where constraints based on the destruction of the star can be imposed in the case of asymmetric non-interacting bosonic dark matter. We should emphasize here that although in the case where WIMPs have self-interactions this maximum WIMPs mass changes, the minimum black hole mass $M_{\rm crit}$ is still valid.  
\begin{table}[ht]
\label{tab:BondiCriticalMasses}
\caption{Critical Mass and $m_{\rm upp}$ in Bondi accretion}
\begin{tabular}{|c|c|c|c|}
\hline
& Free &  Fermi & Boosted\\
\hline
$m_{\rm upp}$ [GeV] & 4.88 & 4.96 & 5.26 \\
\hline
$M$ [GeV] & $1.95\cdot10^{37}$ & $1.91\cdot10^{37}$ & $1.80\cdot10^{37}$  \\
\hline
\end{tabular}
\end{table}
Table II shows the percentage of blocking for the different particles (i.e. quarks, gluons and electrons) for the aforementioned cases for a black hole of mass $M_{\rm crit}$. It is understood that for heavier black holes the blocking will be more significant. It is evident also that the blocking of electron is in all cases insignificant. This is due to the fact that the electron Fermi momentum is much smaller compared to the ones of the quarks.  Once should also notice here that despite the fact that in the boosted blocking case, there is significant blocking of all emitted quarks, the overall effect is rather small i.e. the critical black home mass moves ony about $\sim 10\%$. There are two reasons for this: the first one has to do with the fact that photons, electrons and antiquarks are not blocked at all. The second is due to the fact that $M_{\rm crit}$ for the black hole scales as $f^{1/4}$ (see Eq.~(\ref{evol})). Therefore a significant change of $~40\%$ in $f$ corresponds only to a $~10\%$ in the critical mass. We used a neutron star core density $\rho=6.45 \times 10^{38}~\text{GeV}/\text{cm}^3$ (which is the energy density of the quark system with $\mu=500$ MeV). For the results in Tables I and II we used $c_s=0.17c$. In the case of the maximum possible value of $c_s=c/\sqrt{3}$, one should note that as we mentioned before there is no essential diefference between simple blocking and boosted blocking. Therefore on can easily obtain $M_{\rm crit}$ by simply rescaling the value of $M_{\rm crit}$ from Table I (simple blocking) by a factor of $(0.58/0.17)^{3/4}$. This is because Bondi accretion scales as $\sim c_s^{-3}$ and therefore $M_{\rm crit} \sim c_s^{3/4}$. Therefore in the extreme case of $c_s=0.58c$, $M_{\rm crit}=4.79 \cdot 10^{37}~\text{GeV}$.
\begin{table}[ht]
\label{tab:BondiBlocking}
\caption{Power output blocking percentage at the critical mass for simple Fermi  and boosted Fermi blocking in Bondi accretion.}
\begin{tabular}{|c|c|c|}
\hline
\multirow{2}{*}{} & \multicolumn{2}{c|}{Blocking Percentage} \\
\hline
& {Simple Blocking} & {Boosted  Blocking} \\
\hline
Up Quark & 1.08\% & 55.68\% \\
\hline
Down Quark & 1.14\% & 57.41\% \\
\hline
Strange Quark & 1.02\% & 53.92\% \\
\hline
Electron & $2.4\cdot10^{-5}$\% & $2.3\cdot10^{-4}$\% \\
\hline
Gluon & 56.05\% & 100\%\\
\hline
\end{tabular}
\end{table}

Similarly Tables III and IV show the minimum mass for a growing black hole in the case where accretion is not of Bondi type but of the geometric form we studied in Sec. II. One sees that the critical mass rises with respect to the Bondi acrretion scenario. This is expected since the Bondi accretion rate is larger than the geometric one, thus requiring heavier black holes in order to overcome Hawking radiation.  As in the case of the Bondi accretion scenario, the difference between considering the blocking effect on the emitted modes or not is up to $\sim 13\%$. There is no boosted blocking entry in Table III and IV because the enhanced Fermi momentum of the particles close to the horizon is only an aspect of the Bondi accretion.
\begin{table}[h]
\label{tab:NonBondiCritical}
\caption{Critical mass and $m_{\rm upp}$ for geometric accretion}
\begin{tabular}{|c|c|c|}
\hline
& Free Emission & Fermi Blocking \\
\hline
$m_{\rm upp}$ [GeV] & 1.14 & 1.27  \\
\hline
$M$ [GeV] & $8.33\cdot10^{37}$ & $7.49\cdot10^{37}$ \\
\hline
\end{tabular}
\end{table}

\begin{table}[h]
\label{tab:NonBondiCriticalBlocking}
\caption{Power output blocking percentage at the critical mass for geometric accretion}
\begin{tabular}{|c|c|}
\hline
& Blocking Percentage \\
\hline
Up Quark & 80.58\%  \\
\hline
Down Quark & 81.79\% \\
\hline
Strange Quark & Not Emitted  \\
\hline
Electron & $3.6\cdot10^{-4}$\%  \\
\hline
Gluon & 100\%  \\
\hline
\end{tabular}

\end{table}


The fact that different quarks are blocked to different degrees due to their different Fermi momenta while antiquarks are not blocked can potentially lead to a nonzero net electric charge for the black hole. However charged black holes tend to get rid of the charge by increasing the emission rate of  species with the same charge as that of the black hole and by reducing the rate of emission for the modes with the opposite charge~\cite{Page:1977um}. The calculation we have presented here has been performed assuming neutral black holes. However, if the differences in the blocking among up, down and strange quarks are significant, the black hole will very fast get charged and the emission rates will change accordingly, potentially invalidating our calculation.  In what follows we demonstrate that it is a good approximation in fact to ignore the effect of the charge of the black hole in our cases of interest. 

It was pointed out in~\cite{Page:1977um} that the emission rate of a species that caries charge $z_1e$ out of a black hole of charge $Ze$ is 
\begin{equation}
\frac{dN_q}{dt}=\frac{dN}{dt}e^{4\pi \alpha z_1Z},
\end{equation}
where $\alpha = 1/137$ is the fine structure constant, and $dN/dt$ is the emission rate of the species if the black hole is neutral. Therefore the emission rates of up, down and strange quarks as well as their antiparticles are

\begin{eqnarray}
\frac{dN_u}{dt} = \frac{dN}{dt} \gamma_u e^{\frac{8\pi \alpha Z}{3}}, \qquad  \frac{dN_{\bar{u}}}{dt}&=&\frac{dN}{dt} e^{-\frac{8 \pi \alpha Z}{3}} \\ \nonumber
\frac{dN_d}{dt} = \frac{dN}{dt} \gamma_d e^{-\frac{4\pi \alpha Z}{3}}, \qquad  \frac{dN_{\bar{d}}}{dt}&=&\frac{dN}{dt} e^{\frac{4 \pi \alpha Z}{3}} \\ \nonumber
\frac{dN_s}{dt} = \frac{dN}{dt} \gamma_s e^{-\frac{4\pi \alpha Z}{3}}, \qquad  \frac{dN_{\bar{s}}}{dt}&=&\frac{dN}{dt} e^{\frac{4 \pi \alpha Z}{3}}, \label{charge1}
\end{eqnarray}
where $\gamma_{u,d,s}$ is simply the fraction of the unblocked rate i.e. one minus the fraction of number density that is blocked due to the Fermi surface of the quarks. It is understood that $\gamma=1$ for antiparticles since they are not blocked at all, whereas $\gamma_{u,d,s}$ will be smaller than one. Although the blocking percentages we have presented in Tables II and IV for individual quarks refer to suppression of the power output of the emission, since all quarks have the same peak energy in the spectrum, it is a good approximation to assume that the same suppression holds roughly for the number emission rates. Therefore for example in Bondi accretion with boosted blocking $\gamma_u$ can be read from Table II $\gamma_u=1-0.5568\simeq 0.44$. The accumulation of charge per time in the black hole is
\begin{widetext}
\begin{equation}
\frac{dZ}{dt}=\frac{2}{3} \left ( \frac{dN_{\bar{u}}}{dt}-\frac{dN_u}{dt} \right )+ \frac{1}{3} \left ( \frac{dN_d}{dt} - \frac{dN_{\bar{d}}}{dt} \right )+\frac{1}{3} \left (\frac{dN_s}{dt}-\frac{dN_{\bar{s}}}{dt} \right ).
\end{equation}
\end{widetext}
Using the above and Eq.~(22) we find
\begin{equation}
\frac{dZ}{dt}=\frac{2}{3}\frac{dN}{dt} \left (e^{-\frac{8\pi \alpha Z}{3}}-e^{\frac{4\pi \alpha Z}{3}}+\gamma_{ds}e^{\frac{-4\pi \alpha Z}{3}}-\gamma_u e^{\frac{8\pi \alpha Z}{3}} \right ), \label{zrate}
\end{equation}
where $\gamma_{ds}=(\gamma_d+\gamma_s)/2$. It is easy to show that the solution of the above differential equation with initial condition  $Z=0$ at $t=0$ goes  asymptotically  to a value $Z_0$ which can be found by solving the algebraic equation $dZ/dt=0$. In the case of Bondi accretion with simple and boosted blocking, $Z_0=-7 \times 10^{-5}$, and $Z_0=0.001$ respectively. In the case of geometric accretion with Fermi blocking $Z_0=-0.96$. It is trivial to show that for such low values of $Z$, the effect on the emission due to the accumulated charge is negligible. We found that even in the case of geometric accretion, the effect of the charge can influence the value of the critical mass for the black hole at the order of $1\%$.

In this paper we study how the critical value for the mass of the black hole above which it always grows in the center of a neutron star is influenced by the type of accretion and the fact that degeneracy of quarks at the core of the star can effectively block some modes of Hawking radiation. We find that although quarks can be blocked to great extent, the overall effect and therefore uncertainty in the value of the critical mass is of the order of $\sim 10\%$. On the contrary we find that if accretion does not proceed as a Bondi-like one, the critical black hole mass can be larger by a factor of at most $\sim 4$. \\
 {\bf Acknowledgements:} We would like to thank M. Angeles Perez-Garcia for valuable comments. C.K. is supported by the Danish National Research Foundation, Grant No. DNRF90.

  \end{document}